\documentclass[1 2pt]{article}
\usepackage{amssymb}
\begin{document}
\title{{\bf A 2-Component or  N=2 Supersymmetric Camassa - Holm Equation}}

\author{ Ziemowit Popowicz$^{a}$\\
\\
$^{a}$ Institute of Theoretical Physics\\
University of Wroc\l aw\\
pl. M. Borna 9, 50 -205 Wroc\l aw, Poland\\ }

\maketitle

\begin{center}
{\bf Abstract}
\end{center}
The extended N=2 supersymmetric  Camassa - Holm equation is presented. 
It is accomplished by formulation  the supersymmetric version of  the Fuchssteiner method.
In this framework we use two supersymmetric recursion operators of the N=2 $\alpha = -2,4$ 
Korteweg - de Vries equation and construced two different version of the supersymmetric 
Camassa - Holm equation. The bosonic sector of $N=2$, $\alpha=4$ supersymmetric Camassa - Holm 
equation contains  two component generalization of this equation proposed by Chen, Liu and 
Zhang and as the special case  two component generalized  Hunter - Saxton  equation 
considered by Aratyn, Gomes and Zimerman. As a byproduct of our analysis we defined 
the $N=2$ supersymmetric Hunter - Saxton equation. The bihamiltonian structure is 
constructed for the supersymmetric $N=2$, $\alpha=4$ Camassa - Holm equation.

\newpage

\section*{Introduction}

Camassa and Holm  introduced {\cite{comas}} in 1993  the integrable nonlinear partial differential 
equation 
\begin{equation}
u_t - u_{xxt} = -3uu_x + 2u_xu_{xx} + uu_{xxx} = \big ( uu_{xx} + \frac{1}{2}u_x^2 - \frac{3}{2}u^2\big )_x
\end{equation}
which describes a special approximation of shallow water theory and has been extensively studied 
recently {\cite{constan, chiny, heniek, hone,falqui,fuch}}. It was shown that this equation  possesses the 
bihamiltonian structure, could be solved using  the inverse scattering method and  
has the so called peakons solutions. 

The two component generalization of Camassa - Holm equation 
\begin{eqnarray}
m_t & = & -um_x - 2mu_x + \rho \rho_x \\ \nonumber 
\rho_t & = & - (\rho u)_x, {\label{general}}
\end{eqnarray}
where $m=u-u_{xx}$, has been proposed by Chen, Liu and Zhang  {\cite{chiny}}. This 
generalization, similarly to the Camassa - Holm equation, is 
the first negative flow of the AKNS hierarchy and possesses the interesting peakon and multi - kink 
solutions {\cite{chiny, heniek, falqui}}. Moreover the system (2)  is connected with the 
time dependent Schr{\"{o}}dinger spectral problem {\cite{chiny,ford}}. 
Quite recently Aratyn, Gomes and Zimerman  {\cite{heniek}} showed that the modification of the 
Schr{\"{o}}dinger spectral problem for the equation (2) leads  to  two - component 
Hunter - Saxton   equation {\cite{Saxton}}
\begin{eqnarray}
u_{xxt} &=& -2u_xu_{xx} - uu_{xxx} - \rho\rho_x \\ \nonumber
\rho_t & = & - (\rho u)_x, 
\end{eqnarray}

In this paper we show that both mentioned generalizations  are contained in the bosonic sector of $ N=2 $
extended  supersymmetric version of the Camassa - Holm equation. The idea of using extended supersymmetry 
for the generalization of the soliton equations appeared almost in parallel to the usage of this 
symmetry in the quantum field theory {\cite{wess}}. The main idea of the supersymmetry is to treat boson and 
fermion operators equally. In order to get supersymmetric theory we have to add to a system of $ k $ bosonic 
equations $kN$ fermions and $k(N-1)$ boson fields $(k=1,2,.. N=1,2,..)$ in such a way that the final theory 
becomes supersymmetric invariant. From the soliton point of view we can distinguish two important classes of 
the supersymmetric equations: the non-extended $(N=1)$ and extended $(N>1)$ cases. Consideration of the extended 
case may imply new bosonic equations whose properties need further investigation. This may be viewed as a bonus,
but this extended case is in no way more fundamental than the non-extended one. Interestingly enough, 
some typical supersymmetric effects may occur in the supersymmetrical generalization of the 
soliton theory, compare to the classical case. We mention two of them:
the ambiguity of the roots for the supersymmetric Lax operator {\cite{pop}} and appearance odd of 
Poisson brackets {\cite{pop1}}. 

There are many different methods of supersymmetrization of the classical equations. The most popular 
one is to use the gradation arguments in which to each dependent and independent variables, in a given 
equation,  we associate  some 
weights. We then replace  the bosonic fields by superboson fields. For example,  for the extended $N=2$ 
supersymmetric case  we consider   the supersymmetric analog of dependent variable $u(x,t)$,  which can be thought 
as 
\begin{equation}
\Phi(x,t,\theta_1,\theta_2) = v(x,t) + \theta_1\xi_1(x,t) + \theta_2\xi_2(x,t) + \theta_2\theta_1 u(x,t) 
\end{equation}
where $\theta_1$  and $\theta_2$ are two anticommuting variables while $\xi_1$ and $\xi_2$ are Grassmann valued 
functions and $v$ is an additional "new" bosonic field. In the next step using the usual and 
supersymmetric derivatives 
\begin{equation}
D_1=\partial_{\theta_1} + \theta_1\partial_x, \quad  D_2=\partial_{\theta_2} + \theta_2\partial_x, \quad 
D_1D_2 + D_2D_1 = \partial
\end{equation}
we consider the most general assumption on the  supersymmetric version of the given classical system  in such 
a way that to preserve the given gradation of the supersymmetric equation.  In the last step, we assume,  that 
our supersymmetric generalization should  possesses some special properties as for example,  the existence of 
bihamiltonian structure or Lax operator. 

However this procedure can not be applied to the Camassa - Holm equation,  because this equation does not 
preserve gradation with respect to weight. In order to overcome this problem, we supersymmetrize the Fuchssteiner 
method {\cite{fuch}}, in which the hereditary operator,  responsible for Camassa - Holm hierarchy  is constructed 
out of two different hereditrary recursion operators. We introduce the supersymmetry to the theory, 
considering the supersymmetric analog of these two operators.

The paper is organized as follows. In the  first section 
we summarize  the Fuchssteiner method. In the second section,  using the  supersymmetric recursion 
operators, which creates two different  supersymmetric $N=2$ Korteweg - de Vries equation, we constructed  
two different   supersymmetric  version of Camassa - Holm equation. As a byproduct of our analysis we 
define the extended $N=2$ version of the supersymmetric Hunter - Saxton equation.
In the same section we investigate the 
bosonic sector of this equation. In the last section we describe the bihamiltonian structure of the 
supersymmetric Camassa - Holm equation.

\section*{Camassa - Holm Equation}

In order to construct the Camassa - Holm equation, we briefly describe the method used by 
Fuchssteiner {\cite{fuch}}.
This method based on the following observation. If we have two different hereditary recursion operator 
$R_1$ and $R_2$ and one of them, for example ${R_2}$,  is invertible  then $R=R_1R_2^{-1}$ is also the 
hereditary recursion operator.  Therefore this  operator can be used to construct some new integrable 
hierarchy of equations. Let us present this method for the Camassa - Holm equation. 

First let us consider the hereditary recursion operator for the Korteweg de Vries hierarchy
\begin{equation}
R=c\partial^2 +\lambda(\partial u \partial^{-1} + u).{\label{jeden}}
\end{equation} 
where $c$ and $\lambda$ are an arbitrary constants.
This operator generates the hierarchy of integrable equations in which the first member 
is 
\begin{equation} 
\frac{\partial u}{\partial t} = R u_x = cu_{xxx} + 3\lambda uu_x
\end{equation} 
For $c=1$ and $  \lambda = 2$ we have  famous Korteweg - de Vries equation.

The second recursion operator can be extracted from the first operator   shifting  function  
$ u \rightarrow u + \gamma$ in  recursion operator (6) where $\gamma$ is a constant . 

Indeed after shifting the function $ u $ in (\ref{jeden}) the $R$ operator transforms to 
\begin{equation}
R(u+\gamma) = R_1 + R_2 = (c_1\partial^2 +\lambda(\partial u \partial^{-1} + u)) + 
(c_2\partial^{2} + \lambda\gamma).
\end{equation}
where $c=c_1+c_2$.
It appeared that the recursion operator $ R = R_1R_2^{-1}$ generates new hierarchy of integrable 
equations
\begin{equation}
u_{t_n}=(R_1R_2^{-1})^n u_x
\end{equation} 
Assuming that $u = R_2v = c_2v_{xx} + \lambda\gamma v  $  the first member of the hierarchy is 
\begin{equation}
\lambda \gamma v_t + c_2v_{xxt} = \big ( c_1v_{xx} + \lambda c_2v_{xx}v + \frac{\lambda c_2}{2}v_x^2 + 
\frac{3 \lambda^2\gamma}{2}v^2 \big )_x.
\end{equation}
Now assuming that $ c_1=0,c_2=\gamma=1$ and $ \lambda=-1$ the last equation is exactly the 
Camassa - Holm equation. The second choice $ \gamma = 0, c_1=0 , \lambda = -1$ leads us to the 
Hunter - Saxton equtaion \cite{Saxton}
\begin{equation}
v_t=-(v_{xx}v + \frac{1}{2}v_x^2)_x
\end{equation} 
The bihamiltonian structure of equation (1)  has been constructed  in {\cite{comas}}
\begin{equation}
m_t=J_2\frac{\delta H_1}{\delta m}=\big ( c_1\partial^3 +\lambda(\partial m + m\partial) \big )\frac{\delta H_1}{\delta m}
 = 
J_1\frac{\delta H_2}{\delta m}= \big (c_2\partial^3 +\lambda\gamma\partial) \frac{\delta H_2}{\delta m} 
\end{equation}
where $m= c_2v_{xx} + \lambda\gamma v$ and $H_1= \frac{1}{2} \int dx (c_2v_{xx} + \lambda\gamma v)v ,
H_2 = \frac{1}{2} \int dx \big (2c_1\lambda v_{xx}v^2 + 3c_1v_{xx}v + c_2 \lambda v_x^2 v 
+3\gamma \lambda^2v^3  \big )$. 

As we see the second hamiltonian operator is the same as for the 
Korteweg - de Vries equation and is connected with the Virasoro algebra. For centerless Virasoro 
algebra $c_1=0$ we have additional conserved Hamiltonian $2\int \sqrt m$  which is the Casimir for 
this algebra as well. Using the first Hamiltonian operator to this quantity we obtain the 
Harry Dym type equation of motion
\begin{equation}
m_t=J_1\frac{1}{\sqrt m}
\end{equation}
which reduces to the Harry Dym equation when $c_2=1 , \gamma=0$.

\section*{N=2 Supersymmetric Camassa - Holm and Hunter -Saxton equation} 

We have four different integrable extended $N=2$ supersymmetric extensions of the Korteweg - de Vries equation. 
Three of them 
are connected with the supersymmetric Virasoro algebra \cite{mathie1} while fourth is connected with the 
odd version of supersymmetric Virasoro algebra \cite{pop1}. 
The  first three mentioned   equations are
\begin{eqnarray}
\Phi_t &=& J_2\frac{\delta H_1}{\delta \Phi} = ( D_1D_2\partial + 2\partial \Phi + 2\Phi \partial - D_1\Phi D_1 - 
D_2\Phi D_2 )\frac{\delta H_1}{\delta \Phi}  = \\ \nonumber 
&& \big ( -\Phi_{xx} + 3\Phi(D_1D_2\Phi) + \frac{(\alpha -1)}{2} (D_1D_2\Phi^2) + \alpha \Phi^3\big )_x
\end{eqnarray}  
where $H_1=\frac{1}{2} \int dx d\theta_1 d\theta_2 (\Phi \Phi(D_1D_2\Phi) +\frac{1}{3}\Phi^3)$ and 
parameter $\alpha$ can take three different values $ 1,4,-2 $. This system  has been
extensively studied from different point of view in many papers. 

The properties of these generalizations are 
different for different values of $\alpha$. The Lax operator for $\alpha=4$ has two different roots {\cite{pop}}.
For $\alpha=4$ case instead the bihamiltonian formulation, we deal with the inverse first hamiltonian structure 
{\cite{pop}}, while for  $\alpha=1$ case we have the nonstandard Lax operator {\cite{pop2}} and higher 
order nonlocal recursion opearator {\cite{sorin}}.

The hereditary recursion operator for  $\alpha=4$ and $ \alpha  = -2 $ has been constructed in 
\cite{pop}
\begin{eqnarray}
R_{4} & = &  J_2J_{1,4}^{-1} = cD_1D_2 - \lambda (2\partial \Phi \partial^{-1} -(D_1\Phi)D_1\partial^{-1} -
(D_2\Phi)D_2\partial^{-1} )\\ \nonumber 
R_{(-2)} & = & J_2J_{1,-2} \\ \nonumber 
J_2 & = & cD_1D_2\partial - \lambda (2\partial \Phi + 2 \Phi \partial - D_1 \Phi D_1 - D_2 \Phi D_2 ) \\ \nonumber
J_{1,4} & = & \partial \\ \nonumber
J_{1,(-2)} &=& c D_1D_2\partial^{-1} - \lambda (\partial^{-1}D_1 \Phi D_1 \partial^{-1} + 
\partial^{-1}D_2 \Phi D_2 \partial^{-1} ) 
\end{eqnarray} 
Here $J_2$ is the second  Hamiltonian operator, which is connected with the extended $N=2$ supersymmetric 
version of Virasoro algebra,  $J_{1,4}$ is 
the first hamiltonian operator for  $\alpha = 4$  while  $J_{1,-2}$ is an inverse operator 
of the first Hamiltonian operator for  $\alpha = -2$ case.

In the next we will use these operators in order to adopt the Fuchssteiner method to the 
supersymmetric case. For this pourposes we consider these operatorss independently.

\section*{{\bf A.) $\alpha = 4$}.}

Interestingly the supersymmetric Korteweg - de Vries equation for this value of $\alpha$ is not a first member of 
hierarchy $\Phi_{t_n}= R^n\phi$. The first member is $\Phi_t=((D_1D_2\Phi)+\Phi^2)_x$ while the second 
is our supersymmetric Korteweg - de Vries equation. However let us use the supersymmetric recursion operator $R_4$ 
to the construction of  the supersymmetric analog of the Camassa - Holm recursion operator. In order to find the 
analog of $R_2$ operator let us shift the $\Phi$ superfunction to $\Phi \rightarrow \Phi + \gamma$ in  $R_4$ 
obtaining
\begin{equation}
R_4(\Phi + \gamma)  = R_1 + R_2
\end{equation} 
where 
\begin{eqnarray}
R_{1} & = &  c_1 D_1D_2 - \lambda (2\partial \Phi \partial^{-1} -(D_1\Phi)D_1\partial^{-1} -
(D_2\Phi)D_2\partial^{-1} )\\ \nonumber 
R_2 & = & c_2D_1D_2 -2\lambda \gamma.
\end{eqnarray}
and $c=c_1+c_2$.
Obviously $R_2$ is the hereditary operator and we can consider the hierarchy of equations 
generated by 
\begin{equation}
\Phi_{t_n} = (R_1R_2^{-1})^n \Phi_x
\end{equation}
Assuming that $\Phi=R_2\Upsilon=c_2(D_1D_2\Upsilon) - 2\lambda\gamma\Upsilon$ we obtain that the first 
member in this hierarchy is 
\begin{equation}
c_2(D_1D_2\Upsilon_t) - 2\lambda \gamma \Upsilon_t = c_1(D_1D_2\Upsilon)_x - 2\lambda (\Phi\Upsilon)_x +
\lambda(D_2\Phi)(D_2\Upsilon) + \lambda(D_1\Phi)(D_1\Upsilon) .
\end{equation}
or explicitely as 
\begin{equation}
c_2(D_1D_2\Upsilon_t) - 2\lambda \gamma \Upsilon_t = \big ( c_1(D_1D_2\Upsilon) - 2c_2\Upsilon (D_1D_2\Upsilon)+
4\gamma \lambda^2 \Upsilon^2  + c_2\lambda(D_2\Upsilon)(D_1\Upsilon) \big )_x .
\end{equation}
It is our supersymmetric $N=2$, $\alpha=4$ Comasa - Holm equation.

Let us compute the bosonic sector of the previous equation where all odd function disappear. 
Assuming that $\Upsilon = v + \theta_2\theta_1 u$ in this sector, we obtain 
\begin{eqnarray}
(c_2u-2\gamma\lambda v)_t & = & \big (c_1u+4\gamma\lambda^2 v^2 - 2c_2\lambda vu \big )_x, \\ \nonumber
(-c_2v_{xx} - 2\gamma\lambda u)_t & = & \big ( -c_2\lambda u^2 + 2c_2\lambda v_{xx}v - c_1v_{xx} +
c_2\lambda v_x^2 +8\gamma\lambda^2 vu \big )_x
\end{eqnarray} 
Now introducing new function $ \rho = c_2u-2\gamma\lambda v $ our system  (21) can be rewritten as 
\begin{eqnarray}
\rho_t & =& \frac{1}{c_2}\big (c_1\rho +2c_1\gamma\lambda v -2c_2\lambda\rho v \big )_x \\ \nonumber {\label{krol}}
(-c_2v_{xx} - \frac{4\gamma^2\lambda^2}{c_2} v)_t & = & \big (\frac{2c_1\gamma\lambda}{c_2}\rho +
\frac{4c_1\gamma^2\lambda^2}{c_2}v - c_1v_{xx} + 2c_2\lambda vv_{xx} +\\ \nonumber 
&&  c_2\lambda v_x^2  + \frac{12\gamma^2\lambda^3}{c_2}v^2 - \frac{\lambda}{c_2} \rho^2\big )_x
\end{eqnarray}
Thus we obtained even more general generalization of two - component Camassa - Holm equation then 
equation  (2). 
Interestingly our equation (22)  contains equation (2) as a special case in which 
$c_1=0, c_2^2 =-1, \lambda=\frac{1}{2},\gamma=1 $. 

For $\gamma=c_1=0, \lambda=\frac{1}{2},c_2^2=-1$ our 
system of equation (22) reduces to the two component version of Hunter - Saxton (3).

From that reason the equation (20) when $\lambda=\frac{1}{2}, c_1=0$ 
\begin{equation}
(D_1D_2\Upsilon)_t = \frac{1}{2} \big (  (D_2\Upsilon)(D_1\Upsilon) - 2\Upsilon (D_1D_2\Upsilon) \big )_x .
\end{equation}
could be considered as the $N=2$ supersymmetric Hunter - Saxton equation.

\section*{{\bf B.)  $\alpha = -2$}.}
In contrast to the previous case, now the supersymmetric Korteweg - de Vries equation is a first member of the 
hierarch $\Phi_t=R_{(-2)}\Phi_x$. 
Similarly to the $ \alpha =4 $ case we can  shift the superfunction $\Phi \rightarrow \Phi + \gamma $ in $R_{-2}$
in order to obtain $R_2$ operator.  However then we obtain that  operator $R_2$
contains the $\Phi$ superfunction and therefore  will be not considered here.
The next choice is to use  the $R_2$ operator from  $\alpha=4$ 
case. In this situation we obtain an integro - differential supersymmetric equation and from that reasons we 
will not consider such possibility.

From the other side, we can try to choose  $R_2= \gamma-\partial^2$ operator, the same as in the 
classical situation,  in order to consider the hierarchy of equation generated by 
\begin{equation}
\Phi_t = R_{(-2)}R_2^{-1}\Phi_x
\end{equation}
where now  $\Phi=R_2\Upsilon= \gamma \Upsilon - \Upsilon_{xx}$. 

Assuming that $\lambda=\frac{1}{2}$ and $c=1$  the last equation reads 
\begin{eqnarray}
&& (\gamma\Upsilon - \Upsilon_{xx})_t = \frac{1}{2} \Big (4(D_1D_2\Upsilon)\Upsilon_{xx} - 2\Upsilon_{xx} +
\Upsilon_{xx}\Upsilon_x^2  - 4\gamma(D_1D_2\Upsilon)\Upsilon - \\ \nonumber 
&& \gamma(D_2\Upsilon)(D_1\Upsilon)(D_1D_2\Upsilon)-
 \gamma\Upsilon_{xx}\Upsilon^2 - 
\gamma\Upsilon_x^2\Upsilon - 2\gamma^3\Upsilon^3 + \gamma(D_2\Upsilon)(D_1\Upsilon) \Big )_x + \\ \nonumber 
&& \frac{1}{4} \Big ( (D_2\Upsilon_{xx})(D_2\Upsilon)\Upsilon_{xx} -2(D_2\Upsilon_{xxx})(D_1\Upsilon) - 
  (D_2\Upsilon_{xx})(D_2\Upsilon_x)\Upsilon_x - \\ \nonumber 
&& (D_2\Upsilon_{xx})(D_1\Upsilon_x)(D_1D_2\Upsilon) + (D_2\Upsilon_{xx})(D_1\Upsilon)(D_1D_2\Upsilon_x) - 
(D_2\Upsilon)(D_1\Upsilon_{xx})(D_1D_2\Upsilon_x)  + \\ \nonumber 
&&  
2(D_1\Upsilon_{x})(D_1\Upsilon)\Upsilon_{xxx} +  \gamma\big ((D_2\Upsilon)(D_1\Upsilon))_x(D_1D_2\Upsilon) -
 3\gamma (D_2\Upsilon_x)(D_2\Upsilon)\Upsilon_x + \\ \nonumber 
&& \hspace{3.7cm}   ( D_1 \Rightarrow D_2, D_2 \Rightarrow -D_1) \Big ). 
\end{eqnarray}
It is our supersymmetric $N=2$, $\alpha=-2$ Camassa  - Holm equation.

The bosonic sector in which $\Upsilon = v + \theta_1\theta_2 u$ read 
\begin{equation}
(\gamma v - v_{xx})_t = \frac{1}{2} \Big (4v_{xx}u - 2v_{xx} + v_{xx}v_x^2 -\gamma v_{xx}v^2- 
\gamma v_x^2v +\gamma v^3 -  4\gamma vu \Big )_x,
\end{equation}
\begin{eqnarray}
 && (\gamma u - u_{xx})_t = \frac{1}{2} \Big ( 2u_{xx}u - 2u_{xx}  + u_x^2 - 3\gamma u^2 - 2v_{xxx}v_xu - 
2v_{xxx}v_x -  \\ \nonumber
&& \hspace{2cm}  3v_{xx}^2 + 2v_{xx}v_xu_x - 2\gamma v_{xx}vu  + v_x^2 u_{xx} + \gamma v_x^2u +
\gamma v_x^2 - \\ \nonumber
&& 2v_xvu_x - \gamma v^2u_{xx} + \gamma^2v^2u \Big )_x +  v_{xxxx}v_x + 2v_{xxx}v_{xx}u - 
3\gamma v_{xx}v_xu.
\end{eqnarray}

For  $v=0$ this system reduces to the classical Camassa - Holm equation (1) while for  $v \neq 0$ 
gives us  new generalization of Camassa - Holm  equation.

\section*{{\bf C.)  $\alpha = 1$ }.}

As we mentioned earlier,  the recursion operator for this case, is higher order nonlocal.
From that reason,  the $R_2$ operator obtained shifting the $\Phi$  superfunction,  analogously to the 
previous cases,  leads us to very complicated operator. If we follows in the same way as in $\alpha=4$ case and 
use the the same operator $R_2$ then we obtain very complicated system also. From that reason we will not 
study further this case.

\section*{Bihamiltonian Structure.}

We have constructed this structure  for $\alpha = 4$ case only. 
This structure could  be obtained from the  recursion operator $R_4$ in a similar manner as is  classical 
case {\cite{comas}}. 
In order to end this let us make the following observation. Notice  that the first member in the hierarchy (18) 
could be rewritten as 
\begin{equation}
M_t= J_2(M)J_{1,4}^{-1}R_2^{-1} M_x  = J_2\Upsilon = J_2R_2^{-1} \frac{\delta H_1}{\delta \Upsilon} =
J_2 \frac{\delta H_1}{\delta M}  
\end{equation}
where $M=R_2\Upsilon=c_2(D_1D_2\Upsilon - \lambda\gamma\Upsilon) $ and 
$H_1=\frac{1}{2} \int dx d\theta_1d\theta_2 ((R_2\Upsilon) \Upsilon)$.
As we see this is the second hamiltonian structure and is generated by the same supersymmetric operator
which is responsible for the second  hamiltonian structure of Korteweg - de Vries equation. 

The first hamiltonian structure follows from the following observation. Notice that the equation (20)  
could be rewritten also  as 
\begin{equation}
M_t=\partial \frac{\delta H_2}{\delta \Upsilon} = \partial R_2 R_2^{-1}\frac{\delta H_2}{\delta \Upsilon} =
\partial R_2 \frac{\delta H_2}{\delta M} 
\end{equation}
where 
\begin{eqnarray}
H_2&=&\frac{1}{6} \int dx d\theta_1 d\theta_2 \big (-4c_2\lambda(D_1D_2\Upsilon)(\Upsilon^2 + 
3c_1(D_1D_2\Upsilon)\Upsilon + \\ \nonumber
&& \hspace{3cm} 8\gamma\lambda^2 \Upsilon^3 +  2c_2\lambda(D_2\Upsilon)(D_1\Upsilon) \big )
\end{eqnarray}  
Now the Hamiltonians operators $J_2$ and $(c_2D_1D_2\partial - \lambda\gamma\partial)$ are compatible.  
 We have checked the compatibility using computer algebra Reduce \cite{hearn} and special computer package SUSY2 \cite{susy2}. Therefore we can apply 
these operators to the construction of recursion operator $ J_2\partial^{-1}R_2^{-1}$. Correspondingly this 
operator generates an infinite sequence of conservation laws.

However in this supersymmetric structure 
there is a fundamental difference compare to  the classical situation. The classical second hamiltonian 
operator of the  Camassa - Holm  equation has a Casimir in the form of $\int dx \sqrt m $ and it is 
a Casimnir for centerless  Virasoro algebra also. The extended $N=2$ centerless supersymmetric 
Virasoro algebra does not possesses such 
Casimir  \cite{pop3}. Hence it is impossible to start 
construct  the supersymmetric analog of the classical hierarchy which contains the Harry Dym type 
equation.

\section*{Conclusion.}
In this paper the extended N=2 supersymmetric generalization of Camassa - Holm equation was presented. 
It was accomplished adopting  the Fuchssteiner method of the generation of the Camassa - Holm equation to 
the supersymmetric case. 
In this framework we used two different recursion operators of the N=2 supersymmetric $\alpha = -2,4$ Korteweg - de Vries 
equation and construced two different version of the supersymmetric Camassa - Holm equation. 
The bosonic sector of the $N=2$, $\alpha=4$ supersymmetric Camassa - Holm equation 
contains  two component generalization of this equation proposed by  Chen, Liu and 
Zhang and  two component Hunter - Saxton  equation 
considered by Aratyn, Gomes and Zimerman. As a byproduct of our analysis we defined 
the $N=2$ supersymmetric Hunter - Saxton equation. We have constructed the bihamiltonian structure 
for the supersymmetric $N=2$ , $\alpha = 4 $ Camassa - Holm equation only. For the  $\alpha =-2$ case,  
the supersymmetric Korteweg - de Vries has the inverse first hamiltonian formulation.  From that reasons  
we expect that the same may  occur in the supersymmetric $N=2$, $\alpha =-2$ version of the 
Camassa - Holm equation.  However his point of view needs further investigations.

\end{document}